\RequirePackage[hyphens]{url} 
\documentclass{article}
\usepackage{spconf,amsmath,graphicx, hyperref, multirow, booktabs}

\title{Deep Xi as a Front-End for Robust Automatic Speech Recognition}
\name{Aaron Nicolson and Kuldip K. Paliwal}
\address{Signal Processing Laboratory, Griffith University, Brisbane, Australia}
\begin{document}
%\ninept
%
\maketitle
\begin{abstract}
Current front-ends for robust automatic speech recognition (ASR) include masking- and mapping-based deep learning approaches to speech enhancement. A recently proposed deep learning approach to \textit{a priori} SNR estimation, called Deep Xi, was able to produce enhanced speech at a higher quality and intelligibility than current masking- and mapping-based approaches. Motivated by this, we investigate Deep Xi as a front-end for robust ASR. Deep Xi is evaluated using real-world non-stationary and coloured noise sources at multiple SNR levels. Our experimental investigation shows that Deep Xi as a front-end is able to produce a lower word error rate than recent masking- and mapping-based deep learning front-ends. The results presented in this work show that Deep Xi is a viable front-end, and is able to significantly increase the robustness of an ASR system.
\newline\textbf{Availability}: Deep Xi is available at: \url{https://github.com/anicolson/DeepXi}
\end{abstract}

\section{Introduction}
Recently, an automatic speech recognition (ASR) system was able to achieve human parity on the Switchboard speech recognition task \cite{DBLP:journals/corr/XiongDHSSSYZ16a}. This milestone demonstrates how far ASR research has come in 67 years \cite{OSHAUGHNESSY20082965}. An example of a modern ASR system is Deep Speech \cite{DBLP:journals/corr/HannunCCCDEPSSCN14} --- an end-to-end ASR system that uses a bidirectional recurrent neural network (BRNN) as its acoustic model \cite{650093}. Despite their accuracy on ideal conditions, modern ASR systems are still susceptible to performance degradation when environmental noise is present. A popular method to increase the robustness of an ASR system is to use a front-end to pre-process noise corrupted speech. As the goal of a front-end is to perform noise suppression, a speech enhancement algorithm is typically employed. 

Masking- and mapping-based deep learning approaches to speech enhancement are currently the leading front-ends in the literature \cite{Zhang:2018:DLE:3210369.3178115}. They have the ability to produce enhanced speech that is highly intelligible --- an important attribute for ASR \cite{doi:10.1080/07434619812331278196}. An example of a masking-based approach is the long short-term memory network ideal ratio mask (LSTM-IRM) estimator \cite{doi:10.1121/1.4986931}, which can produce intelligible enhanced speech independent of the speaker. Examples of mapping-based approaches include the fully-connected neural network that employs multi-objective learning and binary mask post-processing to estimate the clean-speech log-power spectra (LPS) (Xu2017) \cite{DBLP:journals/corr/XuDHDL17}, and the time-domain clean-speech estimator that uses encoder-decoder convolutional neural networks (CNNs) as the generator and the discriminator of a generative adversarial network (GAN), called SEGAN \cite{DBLP:journals/corr/PascualBS17}.

Previous generation front-ends include minimum mean-square error (MMSE) approaches to speech enhancement and missing data approaches. Cluster-based reconstruction is a prominent missing data approach, which reconstructs the unreliable spectral components (components with an \textit{a priori} SNR of 0 dB or less \cite{Wang2005}) based on their statistical relationship to the reliable components \cite{RAJ2004275}. MMSE approaches, such as the MMSE short-time spectral amplitude (MMSE-STSA) estimator, rely on the accurate estimation of the \textit{a priori} SNR \cite{1164453}. However, previous \textit{a priori} SNR estimators, like the decision-directed (DD) approach, introduce a tracking delay and a large amount of bias \cite{1164453}.

A deep learning approach to \textit{a priori} SNR estimation was recently proposed, called Deep Xi \cite{NICOLSON201944}. It is able to produce an \textit{a priori} SNR estimate with negligible bias, and exhibits no tracking delay. Deep Xi enabled MMSE approaches to achieve higher quality and intelligibility scores than recent masking- and mapping-based deep learning approaches to speech enhancement. As Deep Xi is able to produce more intelligible enhanced speech than masking- and mapping-based deep learning approaches to speech enhancement, we propose that Deep Xi can further be used as a front-end for robust ASR.

\begin{figure*}[ht!]
% 	\captionsetup{justification=centering}
	\begin{center}
		\includegraphics[scale=0.6]{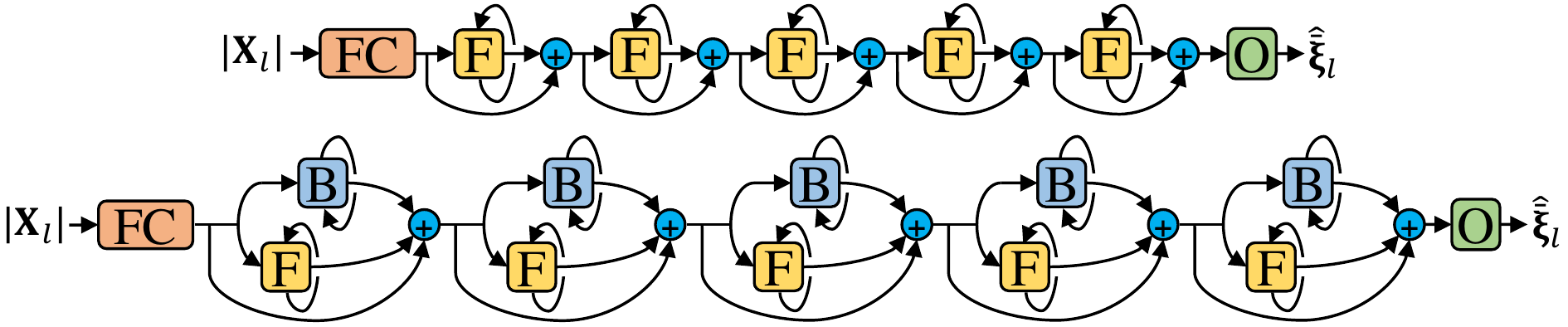}
		\caption{Deep Xi-ResLSTM (top) and Deep Xi-ResBiLSTM (bottom).}
		\label{figd} 
	\end{center}
\end{figure*}

Here, Deep Xi as a front-end is evaluated using real-world non-stationary and coloured noise sources at multiple SNR levels. Deep Speech is used to evaluate each of the front-ends. Deep Speech is suitable for front-end evaluation as it is trained on multiple clean speech corpora and has no implemented front- or back-end techniques. The paper is organised as follows: an overview of Deep Xi is given in Section \ref{secn}; the proposed front-end is presented in Section \ref{seca}; the experiment setup is described in Section \ref{secb}, including a description of each front-end; the results and discussion are presented in Section \ref{secc}; conclusions are drawn in Section \ref{secd}.

\section{Overview of Deep Xi} \label{secn}

In \cite{NICOLSON201944}, the Deep Xi framework for estimating the \textit{a priori} SNR was proposed. The \textit{a priori} SNR is given by:
\begin{equation} \label{equ:a}
\xi[l,k] = \frac{\lambda_s[l,k]}{\lambda_d[l,k]},
\end{equation}
where $l$ is the time-frame index, $k$ is the discrete-frequency index,  $\lambda_s[l,k]=\mathrm{E}\{|S[l,k]|^2\}$ is the variance of the clean speech spectral component, and $\lambda_d[l,k]=\mathrm{E}\{|D[l,k]|^2\}$ is the variance of the noise spectral component. A residual LSTM (ResLSTM) network and a residual bidirectional LSTM (ResBiLSTM) network were used to estimate a mapped version of the \textit{a priori} SNR for each component of the $l^{th}$ time-frame, as shown in Figure \ref{figd}. The mapped \textit{a priori} SNR, $\bar{\pmb{\xi}}_l$, is described in Subsection \ref{sec:b}. The input to each network is the noisy speech magnitude spectrum for the $l^{th}$ time-frame, $|\textbf{X}_l|$. 

The hyperparameters for the ResLSTM and ResBLSTM networks in \cite{NICOLSON201944} were chosen as a compromise between training time, memory usage, and speech enhancement performance. The ResLSTM network, as shown in Figure \ref{figd} (top), consists of five residual blocks, with each block containing an LSTM cell \cite{818041}, $\textbf{F}$, with a cell size of 512. The residual connection is from the input of the residual block to after the LSTM cell activation. $\textbf{FC}$ is a fully-connected layer with 512 nodes that employs layer normalisation followed by ReLU activation \cite{Nair:2010:RLU:3104322.3104425}. The output layer, $\textbf{O}$, is a fully-connected layer with sigmoidal units. The ResBiLSTM network, as shown in Figure \ref{figd} (bottom), is identical to the ResLSTM network, except that the residual blocks include both a forward and backward LSTM cell ($\textbf{F}$ and $\textbf{B}$, respectively) \cite{650093}, each with a cell size of 512. The residual connection is applied from the input of the residual block to after the summation of the forward and backward cell activations. Details about the training strategy for the ResLSTM and ResBiLSTM networks are given in Subsection \ref{sec:a}.

\subsection{Mapped \textit{a priori} SNR training target} \label{sec:b}

The training target for the ResLSTM and ResBiLSTM networks is the mapped \textit{a priori} SNR, as described in \cite{NICOLSON201944}. The mapped \textit{a priori} SNR is a mapped version of the oracle (or instantaneous) \textit{a priori} SNR. For the oracle case, the clean speech and noise in Equation \ref{equ:a} are known completely. This means that $\lambda_s[l,k]$ and $\lambda_d[l,k]$ can be replaced with the squared magnitude of the clean speech and noise spectral components, respectively. In \cite{NICOLSON201944}, the oracle \textit{a priori} SNR (in dB), $\xi_{\rm dB}[l,k]=10\log_{10}(\xi[l,k])$, was mapped to the interval $[0, 1]$ in order to improve the rate of convergence of the used stochastic gradient descent algorithm. The cumulative distribution function (CDF) of $\xi_{\rm dB}[l,k]$ was used as the map. As shown in \cite{NICOLSON201944}, the distribution of $\xi_{\rm dB}$ for a given frequency component follows a normal distribution. It was thus assumed that $\xi_{\rm dB}[l,k]$ is distributed normally with mean $\mu_k$ and variance $\sigma^2_k$: $\xi_{\rm dB}[l,k]\sim\mathcal{N}(\mu_k,\sigma^2_k)$. The map is given by 
\begin{equation} \label{equa}
\bar{\xi}[l,k] = \frac{1}{2}\Bigg[1 + \textrm{erf}\Bigg( \frac{\xi_{\rm dB}[l,k] - \mu_k}{\sigma_k\sqrt{2}} \Bigg) \Bigg],
\end{equation}
where $\bar{\xi}[l,k]$ is the mapped \textit{a priori} SNR. The statistics found in \cite{NICOLSON201944} for each component of $\xi_{\rm dB}[l,k]$ are used in this work. During inference, $\hat{\xi}[l,k]$ is found from $\hat{\xi}_{\text{dB}}[l,k]$ as follows: $\hat{\xi}[l,k] = 10^{(\hat{\xi}_{\text{dB}}[l,k]/10)},$ where the \textit{a priori} SNR estimate in dB is computed from the mapped \textit{a priori} SNR estimate as follows: $\hat{\xi}_{\rm dB}[l,k] = \sigma_k\sqrt{2}\textrm{erf}^{-1}\big(2\hat{\bar{\xi}}[l,k] - 1\big) + \mu_k.$

\begin{figure*}[!ht]
% 	\captionsetup{justification=centering}
	\begin{center}
		\includegraphics[scale=0.65]{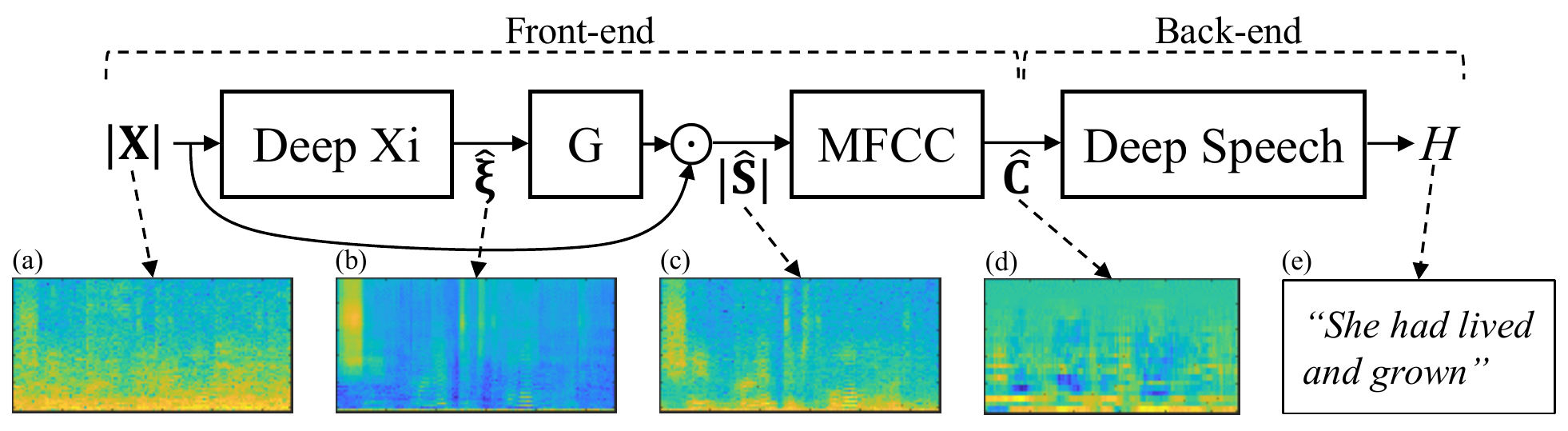}
		\caption{Deep Xi as a front-end for robust ASR. The noisy speech magnitude spectra, $|\textbf{X}|$, is shown in (a). Deep Xi then estimates the \textit{a priori} SNR, $\boldsymbol{\hat{\xi}}$, as shown in (b). An MMSE approach is next used to compute a gain function, $G(\boldsymbol{\hat{\xi}})$, which is then multiplied element-wise with $|\textbf{X}|$ to produce the estimated clean speech magnitude spectra, $|\hat{\textbf{S}}|$, as shown in (c). MFCCs are then computed, producing the estimated clean speech cepstra, $\hat{\textbf{C}}$, as shown in (d). Finally, the hypothesis transcript, $H$, is found using Deep Speech, as shown in (e).}
		\label{figb} 
	\end{center}
\end{figure*}

\section{Deep Xi as a front-end} \label{seca}

In this section, Deep Xi as a front-end for robust ASR is described. Either Deep Xi-ResLSTM or Deep Xi-ResBiLSTM is used to pre-processes the noisy speech before it is given to the back-end of the ASR system. \href{https://github.com/mozilla/DeepSpeech}{Project DeepSpeech}\footnote{Project DeepSpeech is available at: \url{https://github.com/mozilla/DeepSpeech} (model 0.1.1 was used for this research).} is used as the back-end of the ASR system. It is an open source implementation of the Deep Speech ASR system \cite{DBLP:journals/corr/HannunCCCDEPSSCN14}. It uses 26 mel-frequency cepstral coefficients (MFCC) as its input. The process of finding the hypothesis transcription, $H$, is illustrated in Figure \ref{figb}. The process includes the following four steps:
\begin{enumerate}
	\item The \textit{a priori} SNR estimate, $\boldsymbol{\hat{\xi}}$, of the noisy speech magnitude spectra, $|\textbf{X}|$, is found using Deep Xi, $\boldsymbol{\hat{\xi}} = \textrm{Deep Xi}(|\textbf{X}|)$.
	
	\item The estimated clean speech magnitude spectra, $|\hat{\textbf{S}}|$, is found by applying the gain function of an MMSE approach, $G(\cdot)$, to $|\textbf{X}|$: $|\hat{\textbf{S}}|=|\textbf{X}|\odot G(\boldsymbol{\hat{\xi}})$, where $\boldsymbol{\hat{\xi}}$ is used to compute $G(\boldsymbol{\hat{\xi}})$ ($\odot$ denotes the Hadamard product).
	
	\item The input to Deep Speech is the estimated clean speech cepstra, $\hat{\textbf{C}}$, where $\hat{\textbf{C}}$ is computed from $|\hat{\textbf{S}}|$: $\hat{\textbf{C}}=\textrm{MFCC}(|\hat{\textbf{S}}|)$.
	
	\item Deep Speech computes the hypothesised transcript, $H$, from $\hat{\textbf{C}}$: $H=\textrm{Deep Speech}(\hat{\textbf{C}})$.
\end{enumerate}
Deep Xi attempts to match the observed noisy speech to the conditions experienced by Deep Speech during training, i.e. the unobserved clean speech is estimated from the noisy speech. Steps one, two, and three form the front-end of the robust ASR system.

\section{Experiment Setup} \label{secb}

\subsection{Training set}
The \textit{train-clean-100} set from the Librispeech corpus \cite{7178964}, the CSTR VCTK corpus \cite{veaux2017cstr}, and the $si^*$ and $sx^*$ training sets from the TIMIT corpus \cite{garofolo1993darpa} were included in the training set ($74\,250$ clean speech recordings). $5\%$ of the clean speech recordings ($3\,713$) were randomly selected and used as the validation set. The $2\,382$ recordings adopted in \cite{NICOLSON201944} were used here as the noise training set. All clean speech and noise recordings were single-channel, with a sampling frequency of 16 kHz. The noise corruption procedure for the training set is described in the next subsection.

\subsection{Training strategy} \label{sec:a}

Cross-entropy was used as the loss function. The \textit{Adam} algorithm \cite{DBLP:journals/corr/KingmaB14} with default hyperparameters was used for stochastic gradient descent optimisation. A mini-batch size of $10$ noisy speech signals was used. The noisy speech signals were generated as follows: each clean speech recording selected for a mini-batch was mixed with a random section of a randomly selected noise recording at a randomly selected SNR level (-10 to 20 dB, in 1 dB increments).

\subsection{Test set}
Four noise sources were included in the test set: \textit{voice babble}, \textit{F16}, and \textit{factory} from the RSG-10 noise dataset \cite{steeneken1988description} and \textit{street music} (recording no. $26\,270$) from the Urban Sound dataset \cite{Salamon:2014:DTU:2647868.2655045}. 25 clean speech recordings were randomly selected (without replacement) from the \textit{test-clean} set of the Librispeech corpus \cite{7178964} for each of the four noise recordings. To generate the noisy speech, a random section of the noise recording was mixed with the clean speech at the following SNR levels: -5 to 15 dB, in 5 dB increments. This created a test set of 500 noisy speech signals. The noisy speech was single channel, with a sampling frequency of 16 kHz.

%\begin{figure}
%	\captionsetup{justification=centering}
%	\begin{center}
%		\includegraphics[scale=0.75]{./figb}
%		\caption{A section of each noise recording from the test set. \textit{Voice babble} and \textit{street music} are real-world non-stationary noise sources, while \textit{F16} and \textit{factory} are real-world coloured noise sources.}
%		\label{figc} 
%	\end{center}
%\end{figure}

\begin{table*}[ht!]
	\centering
	\footnotesize
	\setlength{\tabcolsep}{3.6pt}
% 	\captionsetup{justification=centering}
	\caption{${\rm WER \%}$ for Deep Speech using each front-end. Real-world non-stationary (\textit{voice babble} and \textit{street music}) and coloured (\textit{F16} and \textit{factory}) noise sources were used. The lowest ${\rm WER \%}$ for each condition is shown in boldface.}

	\begin{tabular}{llllll|lllll|lllll|lllll} 
		\toprule
		\multirow{3}{*}{\bf Method} & \multicolumn{20}{c}{\bf SNR level (dB)}                                                                                                          \\ 
		\cline{2-21}
		& \multicolumn{5}{c|}{\bf Voice babble} & \multicolumn{5}{c|}{\bf Street music} & \multicolumn{5}{c|}{\bf F16} & \multicolumn{5}{c}{\bf Factory}  \\ 
		\cline{2-21}
		& {\bf-5}  & {\bf0}   & {\bf5}   & {\bf10}  & {\bf15}        & {\bf-5}  & {\bf0}   & {\bf5}   & {\bf10}  & {\bf15}        & {\bf-5}  & {\bf0}   & {\bf5}   & {\bf10}  & {\bf15}  & {\bf-5}  & {\bf0}   & {\bf5}   & {\bf10}  & {\bf15}             \\ 
		\hline
		
		Noisy speech & 95.1 & 91.1 & 70.6 & 36.7 & 11.0 & 92.9 & 78.2 & 51.1 & 21.1 & 11.0 & 100.0 & 98.7 & 82.5 & 39.2 & 18.1 & 97.7 & 89.0 & 59.8 & 29.2 & 10.5 \\
		DD & 94.9 & 87.8 & 70.4 & 32.0 & 10.4 & 88.5 & 72.4 & 47.6 & 24.6 & 14.0 & 98.4 & 82.1 & 43.2 & 18.8 & 11.6 & 94.6 & 83.6 & 52.7 & 26.2 & 13.8 \\
		Clust. recon. & 98.2 & 84.5 & 54.2 & 21.0 & 10.2 & 86.2 & 72.2 & 41.3 & 15.4 & 10.1 & 98.0 & 83.1 & 45.4 & 18.6 & 7.0 & 97.1 & 81.6 & 50.3 & 29.2 & 16.7 \\
		Xu2017 & 95.0 & 78.1 & 54.3 & 31.3 & 13.0 & 90.0 & 75.4 & 45.2 & 28.6 & 17.1 & 93.8 & 81.7 & 51.2 & 25.1 & 21.7 & 97.4 & 90.7 & 67.1 & 35.4 & 18.3 \\
		LSTM-IRM & 94.1 & 88.9 & 54.2 & 22.5 & \textbf{9.6} & 89.6 & 67.5 & 39.8 & 16.2 & 7.2 & 97.1 & 79.3 & 45.2 & 21.4 & 12.9 & 94.5 & 73.8 & 43.2 & 18.1 & 10.9 \\
		SEGAN & 95.8 & 79.0 & 44.2 & 19.5 & 10.7 & 84.5 & 58.7 & 32.4 & 12.8 & 11.0 & 94.2 & 76.8 & 45.6 & 19.7 & 7.7 & 91.2 & 70.3 & 45.9 & 18.1 & \textbf{8.5} \\
		Deep Xi-ResLSTM & \textbf{92.9} & 73.3 & 37.1 & 11.7 & 12.0 & 85.5 & 58.1 & 25.2 & 11.5 & \textbf{6.4} & 95.6 & 69.3 & 30.8 & 16.4 & 7.2 & 93.0 & 69.5 & 36.2 & 17.9 & 9.1 \\
		Deep Xi-ResBiLSTM & 95.0 & \textbf{66.0} & \textbf{27.1} & \textbf{10.7} & 10.4 & \textbf{73.0} & \textbf{43.0} & \textbf{23.7} & \textbf{10.3} & 7.0 & \textbf{88.7} & \textbf{56.6} & \textbf{21.4} & \textbf{12.7} & \textbf{4.2} & \textbf{84.9} & \textbf{51.5} & \textbf{29.5} & \textbf{16.8} & 8.8 \\
		
		\bottomrule          
	\end{tabular}
	\label{tabb}
\end{table*}

\subsection{Front-ends}

The configuration of each front-end is described here: 

\textbf{Deep Xi-ResLSTM \& Deep Xi-ResBiLSTM:} The ResLSTM and ResBiLSTM networks are trained using the training set for 10 epochs.

\textbf{SEGAN:} The implementation available at \url{https://github.com/santi-pdp/segan} was used. The available model is retrained using the training set for 50 epochs. 

\textbf{LSTM-IRM:} The LSTM-IRM estimator from \cite{doi:10.1121/1.4986931} is replicated here. The noisy speech magnitude spectrum was used as the input and the training set was used for 10 epochs of training.

\textbf{Xu2017:} The implementation available at: \url{https://github.com/yongxuUSTC/DNN-for-speech-enhancement} is used. 

\textbf{Cluster-based reconstruction:} A diagonal covariance Gaussian mixture model (GMM) consisting of 128 clusters is trained using the k-means++ algorithm, and the expectation-maximisation algorithm \cite{doi:10.1111/j.2517-6161.1977.tb01600.x}. $7\,500$ randomly selected clean speech recordings from the training set are used for training. A BRNN ideal binary mask (IBM) estimator \cite{Nicolson2018}, which is trained for 10 epochs using the training set, is used to identify the unreliable spectral components \cite{RAJ2004275}. 

\textbf{DD:} The MMSE-based noise estimator from \cite{6111268} is used by the DD approach \textit{a priori} SNR estimator \cite{1164453}. The DD approach estimates the \textit{a priori} SNR for the MMSE-STSA estimator \cite{1164453}. 

\subsection{Signal processing}
The following hyperparameters were used to compute the inputs used by the DD approach, cluster-based reconstruction, the LSTM-IRM estimator, Deep Xi-ResLSTM, and Deep Xi-ResBiLSTM. A sampling frequency of 16 kHz was used. The Hamming window function was used for analysis, with a frame length of 32 ms (512 discrete-time samples) and a frame shift of 16 ms (256 discrete-time samples). For each frame of noisy speech, the 257-point single-sided  magnitude spectrum was computed, which included both the DC frequency component and the Nyquist frequency component.

\section{Results and discussion} \label{secc}

% The performance of Deep Xi utilising different MMSE approaches is shown in Table \ref{taba}. The MMSE approaches that were tested included the WF approach, MMSE-STSA estimator, MMSE-LSA estimator, the constrained WF (cWF) approach \cite{Loizou:2013:SET:2484638}, and the square-root WF (SRWF) approach \cite{1455809}. The SRWF approach attained the lowest ${\rm WER\%}$ over all conditions at ${\rm 37.07\%}$, marginaly outperforming the MMSE-STSA estimator (${\rm 37.09\%}$), and the cWF approach ($37.27\%$). Although the cWF approach was not able to achieve the lowest ${\rm WER\%}$ over all conditions, it was able to achieve the lowest ${\rm WER\%}$ for most conditions, especially for \textit{street music}. The performance of the WF approach was the worst amongst the MMSE approaches, with a ${\rm WER\%}$ of $40.03\%$ over all conditions. It did however perform best for two conditions, \textit{voice babble} at -5 dB, and \textit{factory} at 15 dB.

% While the MMSE-STSA estiamtor did not achieve the lowest ${\rm WER\%}$ for many conditions, it was able to achieve a low ${\rm WER\%}$.

In Table \ref{tabb}, Deep Xi-ResLSTM and Deep Xi-ResBiLSTM are compared to both current and previous generation front-ends using Deep Speech. The current front-ends include Xu2017 \cite{DBLP:journals/corr/XuDHDL17}, an LSTM-IRM estimator \cite{doi:10.1121/1.4986931}, and SEGAN \cite{DBLP:journals/corr/PascualBS17}, all of which are either masking- or mapping-based deep learning approaches to speech enhancement. The previous generation front-ends include the DD approach, and cluster-based reconstruction. Here, the square-root WF (SRWF) MMSE approach \cite{1455809} is used by Deep Xi-ResLSTM and Deep Xi-ResBiLSTM, as it attained the lowest word error rate percentage (${\rm WER\%}$) in preliminary testing (over other MMSE approaches such as the MMSE-STSA estimator).

% As seen in Table \ref{tabb}, both Deep Xi-ResLSTM and Deep Xi-ResBiLSTM demonstrated 

% real-world non-stationary and coloured noise sources (unlike)

Deep Xi-ResBiLSTM demonstrates a significant performance improvement, with an average ${\rm WER\%}$ reduction of $9.3\%$ over all conditions when compared to SEGAN. Deep Xi-ResLSTM also demonstrated a high performance, with an average ${\rm WER\%}$ reduction of $3.4\%$ over all conditions when compared to SEGAN. An explanation for the performance of Deep Xi-ResLSTM and Deep Xi-ResBiLSTM as front-ends is given by their speech enhancement performance. In \cite{NICOLSON201944}, Deep Xi-ResLSTM and Deep Xi-ResBiLSTM were both able to produce more intelligible enhanced speech than recent masking- and mapping-based deep learning approaches to speech enhancement. As more intelligible speech improves speech recognition performance \cite{doi:10.1080/07434619812331278196}, Deep Xi-ResLSTM and Deep Xi-ResBiLSTM are both suitable front-ends for ASR. 

SEGAN and the LSTM-IRM estimator both performed poorly at lower SNR levels when compared to the proposed front-ends. Deep Xi-ResBiLSTM performed particularly well at an SNR level of 5 and 10 dB, with an average ${\rm WER \%}$ reduction of $16.9\%$ and $16.6\%$, respectively, over all noise sources when compared to SEGAN. SEGAN and the LSTM-IRM estimator were both able to demonstrate a high performance at an SNR level of 15 dB. However, Deep Xi-ResBiLSTM and Deep Xi-ResLSTM were able to produce an average ${\rm WER \%}$ improvement of $1.9\%$ and $0.8\%$, respectively, over all noise sources at an SNR level of 15 dB when compared to SEGAN. In summary, Deep Xi-ResBiLSTM and Deep Xi-ResLSTM demonstrate a high performance over all SNR levels, especially at low SNR levels.

\section{Conclusion} \label{secd}
In this paper, we investigate the use of Deep Xi as a front-end for robust ASR. Deep Xi was evaluated using both real-world non-stationary and coloured noise sources, at multiple SNR levels. Deep Xi was able to outperform recent front-ends, including masking- and mapping-based deep learning approaches to speech enhancement. The results presented in this work show that Deep Xi is able to significantly increase the robustness of an ASR system.

\bibliographystyle{IEEEbib}
\bibliography{bibliography}

\end{document}